\documentclass[a4paper,11pt]{article}
\usepackage{amssymb}
\usepackage{amsmath}
\usepackage{graphicx}
\usepackage{amsbsy}
\usepackage{xr}
\usepackage{bm}
\usepackage{color}
\usepackage{gensymb}
\usepackage{caption}
\usepackage{subcaption}
\usepackage[space]{grffile}
\definecolor{cream}{RGB}{222,217,201}
\usepackage{siunitx}
\DeclareSIUnit[number-unit-product = {\,}]
\cal{cal}
\DeclareSIUnit\kcal{\kilo\cal}
\DeclareSIUnit\kcal{\kilo\joule\per\mole}
\DeclareSIUnit\molar{\mole\per\cubic\deci\metre}
\DeclareSIUnit\Molar{\textsc{m}}
\usepackage{multirow}

\usepackage{setspace}

\usepackage[utf8]{inputenc}
\usepackage[T1]{fontenc}
\usepackage{authblk}

\usepackage[margin=0.7in]{geometry}

\begin{document}

\title{Marginally compact phase and ordered ground states in a model polymer with side spheres}

\author[1,2]{Tatjana \v{S}krbi\'{c}\thanks{tskrbic@uoregon.edu}}
\author[3]{Trinh Xuan Hoang\thanks{hoang@iop.vast.ac.vn}}
\author[2,4]{Achille Giacometti\thanks{achille.giacometti@unive.it}}
\author[5]{Amos Maritan\thanks{amos.maritan@unipd.it}}
\author[1]{Jayanth R. Banavar\thanks{{\it corresponding author:} banavar@uoregon.edu}}

\affil[1]{\small \it Department of Physics and Institute for Fundamental Science, University of Oregon, Eugene, OR 97403, USA}

\affil[2]{\small \it
Dipartimento di Scienze Molecolari e Nanosistemi,
Universit\`{a} Ca' Foscari di Venezia
Campus Scientifico, Edificio Alfa,
via Torino 155, 30170 Venezia Mestre, Italy}

\affil[3]{\small \it Institute of Physics, 
Vietnam Academy of Science and Technology,
10 Dao Tan, Ba Dinh, Hanoi 11108, Vietnam}

\affil[4]{\small \it European Center for Living Technologies (ECLT),
Ca' Bottacin, Dorsoduro 3911 Calle Crosera, 30123 Venezia, Italy}

\affil[5]{\small \it Dipartimento di Fisica e Astronomia,
Universit\`{a} di Padova and INFN
via Marzolo 8, 35131 Padova, Italy}

\date{}

\maketitle

\begin{abstract}
We present the results of a quantitative study of the phase behavior of a model polymer chain with side spheres using two independent computer simulation techniques. We find that the mere addition of side spheres results in key modifications of standard polymer behavior. One obtains a novel marginally compact phase at low temperatures, the structures in this phase are reduced in dimensionality and are ordered, they include strands assembled into sheets and a variety of helices, and at least one of the transitions on lowering the temperature to access these ordered states is found to be first order. Our model serves to partially bridge conventional polymer phases with biomolecular phases.
\end{abstract}

\newpage

Our principal goal is a careful quantitative computational analysis of a simple chain model of tethered spheres with side spheres attached to the main chain spheres.  Our motivation for attaching side spheres comes from proteins, the amazing molecular machines of life. The side spheres merely restrict the conformational space that the chain can explore. In spite of this somewhat innocuous role, we demonstrate that there are significant and surprising effects on the phase diagram and the nature of the ground state conformations. We alert the reader that our paper is not about proteins but rather about a simple standard model of polymer physics.  

Some small globular proteins fold in an all-or-nothing manner under physiological conditions -- the transition from the unfolded to a folded state is akin to a first order phase transition albeit for a finite size system \cite{globule_first_order}. The folding is rapid as well as reproducible in that a given protein folds into the same native state structure upon unfolding and refolding. The native state structures of proteins are modular and are built up of building blocks of helices and zig-zag strands assembled into almost planar sheets \cite{lesk_bahar_books}. The simplest model of a homopolymer (made up of just one type of monomer) chain of tethered spheres does $\it not$ account for many of these observations. For a generic range of attractive interaction, mimicking the mediating influence of the water, we obtain a ``continuous" transition from a high temperature expanded coil phase to a  globular phase at the so-called $\theta$-temperature \cite{degennes_lifshitz_globules} via an ideal coil state. On lowering the temperature further, we find indications of a second transition in our computer simulations into a more compact unstructured globular phase. There are no building blocks of helices or sheets nor is there any evidence of crystalline ordering, e.g. tethers passing through a face-centered cubic crystalline arrangement (fcc) \cite{Chaikin2000}. There is no rapid, let alone reproducible, folding into a specific ground state. Motivated by proteins, our principal goal here is to quantitatively study the effects of modifying the standard homopolymer chain model in the simplest possible way by adding side spheres to the main chain spheres.

Proteins are made up of 20  types of naturally occurring amino acids, each with a distinctive side chain. Incorporating this feature would make the problem inherently complex because now the details of the distinct monomer types along with their interactions become a necessary part of the story. Here instead we build a bridge between the homopolymer and protein behaviors by incorporating a simple attribute of proteins into the polymer model. We seek to study the impact of side spheres (attached to all but the two end main spheres) sticking out in the negative normal direction in the local Frenet coordinate frame \cite{kamien_RevModPhys}. We find a marginally compact phase at low temperatures between the compact globular phase and a less well-packed restricted coil phase upon varying the side sphere size. The marginally compact phase is the analog (on the side sphere size axis not the temperature axis) , for chain molecules, of the liquid crystal phase \cite{Chaikin2000}, which is a sensitive phase of matter with applications in displays, mood rings, and sensors. The vicinity of the marginally compact phase to other phases confers sensitivity to a chain poised in it. The structures are ordered and the transition to the marginally compact phase upon lowering the temperature is first order. This is not unexpected because the symmetry of the unstructured coil phase is distinct from that of the ordered marginally compact phase \cite{Chaikin2000}. We find novel ground state structures including a planar sheet made up of zig-zag strands along with a variety of helices and dual helices. We find, generally, that the limitations imposed by our study of moderate length chains as well as difficulties associated with lack of equilibration are greatly reduced as the side sphere size increases. Even though the simple addition of side spheres already introduces many protein-like features, we find clear differences between the marginally compact structures and protein native state structures underscoring that other essential features yet need to be incorporated into simple chain models to describe the amazing molecular machines of life. 

The tangent homopolymer chain comprises $N$ spherical beads of diameter $\sigma$ (set equal to 1) tethered into a chain. Consecutive beads are kept at a fixed distance $b=\sigma$ and non-consecutive beads are not allowed to overlap resulting in excluded volume \cite{polymer_books}. Solvent effects are incorporated by including a short range square-well attraction between beads separated in sequence by at least three. The range of attraction of the square well, $R_c/\sigma$, measured in units of the sphere diameter is a free parameter. 

The phase diagram as a function of the reduced temperature (measured in units of the attractive potential, which can also be set equal to $1$) has been studied over the years by many different groups using various methods \cite{binder_karplus_globules}. It has a high temperature swollen phase also called the coil phase, where the chain is in a relatively open stretched conformation,  which is dominated by entropy. We find two ``continuous" transitions upon lowering the temperature. The first is a transition to a collapsed globule phase at the so-called $\theta$--temperature, and the second is to an unstructured compact globule phase.  At low temperatures, the minimization of the energy or, equivalently, the maximization of the number of attractive contacts is the dominating factor resulting in the compact globule phase (for a chain of moderate length $\approx 100$) with many degenerate conformations having no discernible order and a high number of attractive contacts. There are two unresolved issues of importance: first, what might happen for a much larger system -- would a highly compact crystalline arrangement (such as Hamiltonian walks on a fcc lattice be the true ground states) and second, could the tethering of the spheres thwart equilibration -- are we observing glassy behavior characterized by low-lying metastable minima? Indeed, there is some evidence in careful simulations \cite{binder_karplus_globules} suggesting some murkiness in reliably identifying the ground state. 

We generalize the simple polymer model by mounting on each main chain sphere (except the first and last) tangent side spheres in the negative normal direction. The side spheres provide steric constraints both on main chain spheres and other side spheres and reduce the conformational space. In our analysis, the side spheres do not interact except that they are not allowed to overlap with each other and with the main chain spheres. A partial penetrability would result in an effective smaller side sphere size but would not change the results qualitatively. The side sphere diameter, measured in units of the main chain sphere diameter $\sigma_{sc}/\sigma$, is the second and final parameter of the model. We do not impose any attraction between beads separated in sequence by two in order to not suppress the bond bending angles $\theta$ artificially. The conformation of a chain of constant bond length is fully specified by two angle variables at each location \cite{protein_science_2021}, the bond bending angle $\theta$ and the torsion angle $\mu$, the angle between successive binormals of a Frenet local coordinate system. 

There have been previous insightful studies of heteropolymer lattice protein models with side chains (in contrast, our work here deals with an off-lattice homopolymer), which have noted enhanced cooperative folding \cite{lattice_protein_models} due to the denser packing in the interior of the structure. These studies did not fix the side chain location in the Frenet coordinate system of the main chain residues like we do here. Our own earlier studies \cite{our_side_chain_papers} of homopolymers with side spheres mostly dealt with overlapping main chain spheres (the overlap was overtly introduced to replace the spurious spherical symmetry associated with individual main chain spheres with an axial symmetry befitting a chain). Furthermore, in the earlier work, the attractive potential was extended to pairs of main chain spheres separated by $2$ along the chain (unlike $3$ in our studies here), which promoted the occurrence of artificially small bending angles $\theta$. Finally, the machinery for the quantitative characterization of the variety of helical geometries observed in the simulations using the ($\theta$,$\mu$) variables is being used here for the first time.

Our Monte-Carlo simulations were carried out using two complementary methods: microcanonical Wang-Landau (WL) simulations \cite{wang_landau} and replica exchange (RE) (or parallel tempering) canonical simulations \cite{swendsen_wang}. WL entails the filling of consecutive energy histograms to derive the density of states $g(E)$. The acceptance probability is chosen to promote moves exploring less populated energy states seeking to flatten energy histograms over the course of the simulations thus finding lower energy states. In all cases, 28--30 levels of iterations were carried out with a flatness criterion in each iteration of at least 80\%, ensuring convergence of the $g(E)$ allowing for thermodynamic quantities to be calculated. The RE approach entails running canonical simulations in parallel at M different temperatures, $T_i$, $i = 1, 2,\cdots M$. Each simulation can be thought of as a replica, or a system copy in thermal equilibrium. The key advantage is the possibility of swapping replicas at different temperatures without affecting the ``equilibrium" condition at each temperature, permitting rapid equilibration even when there is a rugged landscape. Both methods employed standard local moves including crankshaft, reptation and end-point along with the non-local pivot move. The results using both methods were completely consistent with each other and each method was useful to benchmark the other.

We now proceed to discuss our results. Figure 1 shows the distinct characteristics of chain conformations in the compact globular (zero temperature possibly ``non-equilibrium" phase maximizing attractive interactions while respecting steric constraints) phase, the coil (infinite temperature--high entropy) phase, and a restricted coil phase in ($\theta$,$\mu$) cross-plots. Panels a) and b) are for chains with no side spheres whereas the chain in panel c) has large side spheres with $\sigma_{SC}/\sigma = 2.8$. There is a tendency, in the ground state, for a chain to bend maximally (favor the smallest sterically permitted value of $\theta$) and avoid planar conformations (by exhibiting a local minimum at $\mu=180^{\circ}$ in contrast to a local maximum in the coil phase). 

We monitored the specific heat as a function of temperature for three different chain lengths in order to understand the nature of all the phase transition(s). As an example, the left panel of Figure 2 depicts the specific heat versus temperature plot for $R_c/\sigma = 1.6$ in the absence of side spheres with the inset showing the canonical energy probability distributions at three temperatures in the vicinity of the lower temperature ``continuous" transition. Our studies become more reliable and the transition becomes first order on introducing large enough side spheres. The middle panel shows the temperature dependence of the specific heat for $\sigma_{SC}/\sigma = 1$, whereas the right panel is for $\sigma_{SC}/\sigma = 1.5$. The ground state for the former case is a precessing helix whereas the ground state for the latter case is a structured dual helix (see third and fourth panels of Figure 3). For the chain lengths we have studied and likely in the infinite chain length limit, the presence of side spheres greater than or equal to $\sigma_{SC}/\sigma = 0.9$ does not allow for folding of the helix or the dual helix onto itself to create a hairpin structure and avail of more attractive main sphere interactions. The scaling of the specific heat peak with system size and the two-peaked structure in the energy probability distribution in the vicinity of the transition to the ground state are both signatures of a first order transition, albeit for a finite size chain. Note that the first order transition is weakened upon increasing the side sphere size from 1 to 1.5 (Figure 2 insets).

Figure 4 shows the phase diagram in the temperature-side sphere size plane for a fixed range of attractive interaction $R_c/\sigma = 1.6$. The region of the phase diagram for side sphere size less than 0.9 seems robust (independent of Monte Carlo technique and different runs), yet is possibly unreliable because of the glassy behaviour and finite size effects (see shaded region of Figure 4). One observes both continuous (blue-diamond) as well as first order (red-circle) transitions. For no side spheres, one observes two continuous phase transitions. Around a value of $\sigma_{SC}/\sigma = 0.6$, there is a new feature in the ground state with the unstructured compact globule giving way to an ordered ground state. The higher transition is still a continuous coil-unstructured globule transition, whereas the lower transition between the globule and the structured ground state is first order because of the distinct symmetries of the two phases. This situation persists until $\sigma_{SC}/\sigma = 0.8$, where there is a steep increase in the coil-globule transition temperature along with changes in the nature of the structured ground state. Starting at $\sigma_{SC}/\sigma = 0.9$, the results of the simulations become reliable. The globule is replaced by a structured dual helix at intermediate temperatures, and the ground state becomes a precessing helix (see the third panel in Figure 3). The lower transition temperature diminishes to zero on increasing the side sphere size and there is just a single coil-double helix weakening first order transition for $\sigma_{SC}/\sigma$ greater than 1.4 (Figure 2). The transition temperature continues to decrease as the side sphere size increases -- attractive contacts are harder to make in any sustained manner. Ultimately, for large enough side sphere size, even at low temperatures, one obtains a coil phase with a greatly reduced phase space because of the steric constraints imposed by the large side spheres (Figure 1 c). 

Figures 3 and 5 show some of the ordered ground state structures in the marginally compact phase along with their ($\theta$,$\mu$) plots. Even the small sampling of structures shown here present a beauty and richness not previously observed in polymer models or for that matter in standard biological systems. The idealized structures (whose topology of attractive contacts are faithfully realized in computer simulations albeit with some variations) include a planar sheet of zig-zag strands, a variety of distinct precessing helices (with repeating ($\theta,\mu$) angles), a dual helix, a helix and a dual helix with a straight chain segment (rod) penetrating it. The precessing helices are in fact two or three uniform helices intercalated between each other. Our quantitative analysis is facilitated by the ($\theta,\mu$) description of the chain conformations.

We find that one obtains a compact globule to coil transition in two distinct ways: first, for the standard case of no side spheres, the globule to coil continuous transition occurs on increasing the temperature;  second, at zero temperature on increasing the size of the side spheres, the globule phase switches to the coil phase through a sequence of first order transitions passing through structured ground states that we denote as marginally compact states. They are marginally compact because they are more compact than the coil phase structures but less so than the globular compact phase structures.

A tangent chain of spheres has problems associated with spurious symmetries. A spherical monomer looks the same when viewed from any direction. However, when strung along a chain, then this spherical symmetry becomes spurious. A chain has a tangent direction at each location and thus the isotropy is lost. Attaching side spheres in the negative normal direction is one way to overtly move away from this fake isotropy of the constituent spheres. One might imagine the compact globule and the somewhat open entropically favored coil states as analogs of the crystal and liquid phases. The marginally compact phase then is the analog of the liquid crystal phase \cite{Chaikin2000}. First, the marginally compact structures lie in the vicinity of transitions to other phases and, second, they arise on breaking the spurious symmetry in the model associated with spherical monomers constituting a uniaxial chain. Liquid crystals are very sensitive to the right types of perturbations and one would expect that to hold as well for the structures in the marginally compact phase. Second, the marginally compact conformation has a Goldilocks compactness between that of a dense globule (in which the conformation seeks to maximize the number of attractive contacts somewhat akin to a sphere surrounded by a dozen spheres in a face-centered-cubic lattice) and a coil conformation (in which a few chance contacts are made but the conformation is a typical self-avoiding walk). Finally, the ordered marginally compact structures are all reduced dimensional structures (topologically one dimensional helices or two dimensional sheets), just as liquid crystals are not ordered in the same manner in all three directions. 

A marginally compact conformation has a Goldilocks compactness between that of a dense globule (in which the conformation seeks to maximize the number of attractive contacts somewhat akin to a sphere surrounded by a dozen spheres in a face-centered-cubic lattice) and a coil conformation (in which a few chance contacts are made but the conformation is a typical self-avoiding walk). Striking examples of marginally compact conformations presented in our paper are dimensionally reduced conformations such as a two dimensional sheet and one dimensional helices and dual helices. Interestingly, these conformations tend to be ordered promoting a first order phase transition between a disordered higher temperature phase and this ordered marginally compact phase. 

We have presented quantitative studies of the phase behavior of a tangent polymer chain with side spheres. Our results demonstrate that the mere addition of side spheres results in key modifications of the model behavior. As is observed in our simulations, proteins \cite{lesk_bahar_books} fold in an all or nothing fashion akin to the behavior at a first order transition, their structures are made up of reduced dimensional building blocks including topologically one dimensional helices and two dimensional sheets. Proteins are stable yet sensitive and are able to carry out a dizzying array of functions. In this regard, they are reminiscent of liquid crystals. Despite the encouraging results reported here, the simple model we have studied is still missing some essential ingredients for faithfully describing proteins. The symptoms of these missing elements are that the structures we obtain are not those found in proteins. Unlike the rich variety of helices found here, the $\alpha$-helix is characterized by ($\theta,\mu$) values around ($91.3 \pm 2.2^{\circ}$,$49.7 \pm 3.9^{\circ}$) \cite{physics_meets_biology}, which we do not find in our studies here. Also, the sheet structure in proteins is qualitatively different from that found in our model -- the key difference is that adjacent strands in proteins are in phase with each other unlike the out of phase packing, which maximizes the number of pair-wise contacts here. Third, we do not observe coexistence of helices and sheets in our model here which is a crucial ingredient for obtaining the diversity of protein folds. A powerful hint regarding the missing ingredient comes from recent work \cite{physics_meets_biology} that used a tube-like description to derive the building blocks of protein structures from first principles with no adjustable parameters. We look forward to studies incorporating a tube description along with side chains to further understand proteins in a simplified manner. More generally, our work here ought to be of relevance for understanding the phase behavior of polymers \cite{polymer_books}. 

{\bf Acknowledgements:} We are indebted to Pete von Hippel for his warm hospitality. {\bf Funding:} This project received funding from the European Union’s Horizon 2020 research and innovation program under the Marie Sk\l odowska-Curie Grant Agreement No 894784. The contents reflect only the authors’ view and not the views of the European Commission. Support from the University of Oregon (through a Knight Chair to JRB), Vietnam National Foundation for Science and Technology Development (NAFOSTED) under grant number 103.01-2019.363 (TXH), University of Padova through ``Excellence Project 2018'' of the Cariparo foundation (AM), MIUR PRIN-COFIN2017 Soft Adaptive Networks grant 2017Z55KCW and COST action CA17139 (AG) is gratefully acknowledged. The computer calculations were performed on the Talapas cluster at the University of Oregon.
{\bf Author contributions:} T\v{S} carried out the calculations with significant help from TXH and under the guidance of JRB. JRB wrote the paper. All authors helped understand the results and reviewed the manuscript.
{\bf Conflict of interest}: The authors declare that they have no conflict of interest.


\newpage



\begin{thebibliography}{999}

\bibitem{globule_first_order}
C.~B.~Anfinsen and H.~A.~Scheraga, ``Experimental and theoretical aspects of protein folding''.
{\em Adv. Prot. Chem.} {\bf 29}, 205--300 (1975); 
R.~Zwanzig, ``Two-state models of protein folding kinetics''. 
{\em Proc. Natl. Acad. Sci. USA} {\bf 94}, 148--150 (1997);
C.~Maffi, M.~Baiesi, L.~Casetti, F.~Piazza, and P.~De Los Rios,
``First-order coil-globule transition driven by vibrational entropy''.
{\em Nat. Commun.} {\bf 3}, 1065 (2012);  A.~Badasyan et al., ``The finite size effects and two-state paradigm of protein folding''.
{\em Int. J. Mol. Sci.} {\bf 22}, 2184 (2021).

\bibitem{lesk_bahar_books}
A.~M.~Lesk, {\em Introduction to Protein Science: Architecture, function and genomics}.
\newblock{Oxford University Press, 2004};
I.~Bahar, R.~L.~Jernigan and K.~A.~Dill, {\em Protein Actions}.
\newblock{Garland Science, Taylor \& Francis Group, 2017}.

\bibitem{degennes_lifshitz_globules}
P.~G.~de~Gennes, ``Collapse of a polymer chain in poor solvents''.
{\em J. Physique Lett.} {\bf 36}, 55--57 (1975);		
I.~M.~Lifshitz, Y.~A.~Grosberg, A.~R.~Khokhlov,
``Some problems of the statistical physics of polymer chains with volume interaction''.
{\em Rev. Mod. Phys.} {\bf 50}, 683--713 (1978).
		
\bibitem{Chaikin2000}
P.~Chaikin and T.~Lubensky, {\em Principles of Condensed Matter Physics}.
\newblock Cambridge University Press, 2000.

\bibitem{kamien_RevModPhys}
R.~D.~Kamien, ``The geometry of soft materials: a primer''.
{\em Rev. Mod. Phys.} {\bf 74}, 953--971 (2002).

\bibitem{polymer_books}
P.~G.~de~Gennes. {\em Scaling Concepts in Polymer Physics}.
\newblock{Cornell University Press, 1979};
A.~R.~Khokhlov, A.~Y.~Grosberg and V.~S.~Pande, {\em Statistical Physics of Macromolecules
(Polymers and Complex Materials)};
\newblock{American Institute of Physics, 1994}.
M.~Rubinstein and R.~H.~Colby, {\em Polymer Physics (Chemistry)}.
\newblock{Oxford University Press, 2003}.

\bibitem{binder_karplus_globules}
Y.~Zhou, M.~Karplus and J.~M.~Wichert and C.~K.~Hall,
``Equilibrium thermodynamics of homopolymers and clusters: Molecular dynamics and Monte Carlo simulations of
systems with square-well interactions'',
{\em J. Chem. Phys.} {\bf 107}, 10691--10708 (1997);
M.~P.~Taylor, W.~Paul and K.~Binder, ``Phase transitions of a single polymer chain: A Wang-Landau simulation study''.
{\em J. Chem. Phys.} {\bf 131}, 114907 (2009);
M.~P.~Taylor, W.~Paul and K.~Binder, ``All-or-none proteinlike folding transition of a flexible homopolymer chain''.
{\em Phys. Rev. E} {\bf 79}, 050801 (2009);
M.~P.~Taylor, W.~Paul and K.~Binder, ``Applications of the Wang--Landau Algorithm to Phase Transitions
of a Single Polymer Chain''.
{\em Polymer Science} {\bf 55}, 23--38 (2013);
M.~P.~Taylor, W.~Paul and K.~Binder,
``On the polymer physics origins of protein folding thermodynamics'',
{\em J. Chem. Phys.} {\bf 145}, 174903 (2016).

\bibitem{protein_science_2021}
T.~\v{S}krbi\'{c}, A.~Maritan, A.~Giacometti and J.~R.~Banavar,
``Local sequence-structure relationships in proteins''.
{\em Protein Sci.} (2021), DOI: 10.1002/pro.4032.

\bibitem{lattice_protein_models}
A.~Kolinski and J.~Skolnick, ``Discretized model of proteins. I. Monte Carlo study of cooperativity inhomopolypeptides''.
{\em J. Chem. Phys.} {\bf 97}, 9412--9426 (1992);
M.~H.~Hao and H.~A.~Scheraga,
``Monte Carlo Simulation of a First-Order Transition for Protein Folding''.
{\em J. Phys. Chem.} {\bf 98}, 4940--4948 (1994);
M.~H.~Hao and H.~A.~Scheraga,
``Statistical Thermodynamics of Protein Folding: Sequence Dependence''.
{\em J. Phys. Chem.} {\bf 98}, 9882--9893 (1994);
A.~Kolinski, W.~Galazka and J.~Skolnick,
``On the Origin of the Cooperativity of Protein Folding: Implications From Model Simulations''.
{\em Proteins} {\bf 26}, 271--287 (1996);
D.~K.~Klimov and D.~Thirumalai,
``Cooperativity in protein folding: from lattice models with side chains to real proteins''.
{\em Folding \& Design} {\bf 3}, 127--139 (1998); V.~V.~Vasilevskaya, P.~G.~Khalatur and A.~R.~Khokhlov,``Conformational Polymorphism of Amphiphilic Polymers in a Poor Solvent''. {\em Macromolecules} {\bf 36}, 10103--10111 (2003). 


\bibitem{our_side_chain_papers}
J.~R.~Banavar, M.~Cieplak, T.~X.~Hoang, and A.~Maritan,
``First-principles design of nanomachines''.
{\em Proc. Natl. Acad. Sci. USA} {\bf 106}, 6900--6903 (2009);
T.~\v{S}krbi\'{c}, A.~Badasyan, T.~X.~Hoang, R.~Podgornik and A.~Giacometti,
``From polymers to proteins: the effect of side chains and broken symmetry on the formation of secondary structures within a Wang-Landau approach''.
{\em Soft Matter} {\bf 12}, 4783--4793 (2016);
T.~\v{S}krbi\'{c}, T.~X.~Hoang and A.~Giacometti,
``Effective stiffness and formation of secondary structures in a protein-like model''.
{\em J. Chem. Phys.} {\bf 145}, 084904 (2016);
T.~\v{S}krbi\'{c}, T.~X.~Hoang, A.~Maritan, J.~R.~Banavar and A.~Giacometti,
``The elixir phase of chain molecules''.
{\em Proteins} {\bf 87}, 174--175 (2019);
T.~\v{S}krbi\'{c}, T.~X.~Hoang, A.~Maritan, J.~R.~Banavar and A.~Giacometti,
``Local symmetry determines the phases of linear chains: a simple model for the self-assembly of peptides''.
{\em Soft Matter} {\bf 15}, 5596--5613 (2019);
T.~\v{S}krbi\'{c}, J.~R.~Banavar and A.~Giacometti,
``Chain stiffness bridges conventional polymer and bio-molecular phases''.
{\em J. Chem. Phys.} {\bf 151}, 174901 (2019).

\bibitem{wang_landau}
F.~Wang and D.~P.~Landau,
``Efficient, Multiple-Range Random Walk Algorithm to Calculate the Density of States''.
{\em Phys. Rev. Lett.} {\bf 86}, 2050--2053 (2001).

\bibitem{swendsen_wang}
R.~H.~Swendsen and J.~S.~Wang,
``Replica Monte Carlo Simulation of Spin-Glasses''.
{\em Phys. Rev. Lett.} {\bf 57}, 2607--2609 (1986).

\bibitem{physics_meets_biology}
T.~\v{S}krbi\'{c}, A.~Maritan, A.~Giacometti, G.~D.~Rose and J.~R.~Banavar,
``Building blocks of protein structures -- physics meets biology'', 
in press in {\em Phys. Rev. E}; bioRxiv DOI: 10.1101/2020.11.10.375105

\end{thebibliography}

\newpage

\begin{figure}[htpb]
\centering
\captionsetup{justification=raggedright,width=\linewidth}
\includegraphics[width=0.7\linewidth]{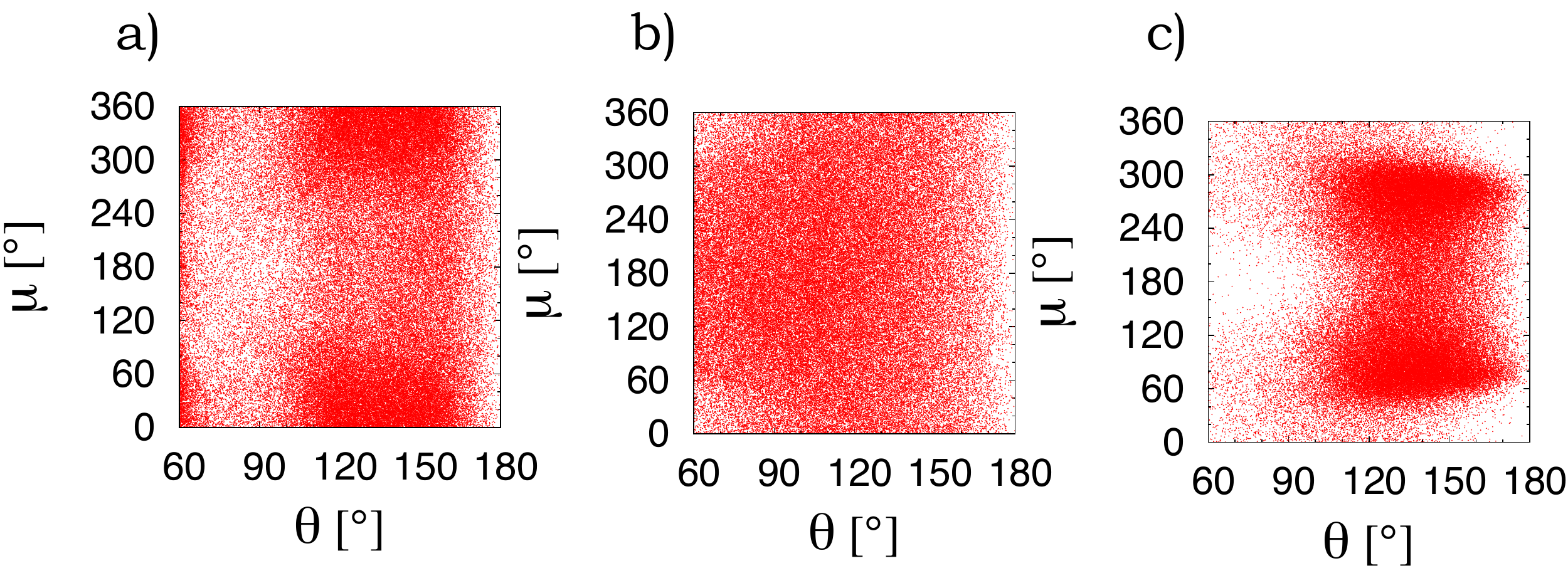}
	\caption{($\theta$,$\mu$) cross-plots showing local structure in three unstructured phases of a standard polymer chain. $\theta$ is the bond-bending angle and $\mu$ is the torsion angle. Panel a) shows the the low temperature compact globular phase in the absence of side spheres for a chain of length $N=60$. Panels b) and c) show the infinite temperature coil phase, but this time for $N=20$. Panel b) is for a chain with no side spheres whereas panel c) is for a chain with large side spheres of size $\sigma_{SC}/\sigma = 2.8$. All three panels show 100,000 points. 
\label{fig:Fig_1}}
\end{figure}

\newpage

\begin{figure}[htpb]
\centering
\captionsetup{justification=raggedright,width=\linewidth}
\includegraphics[width=0.7 \linewidth]{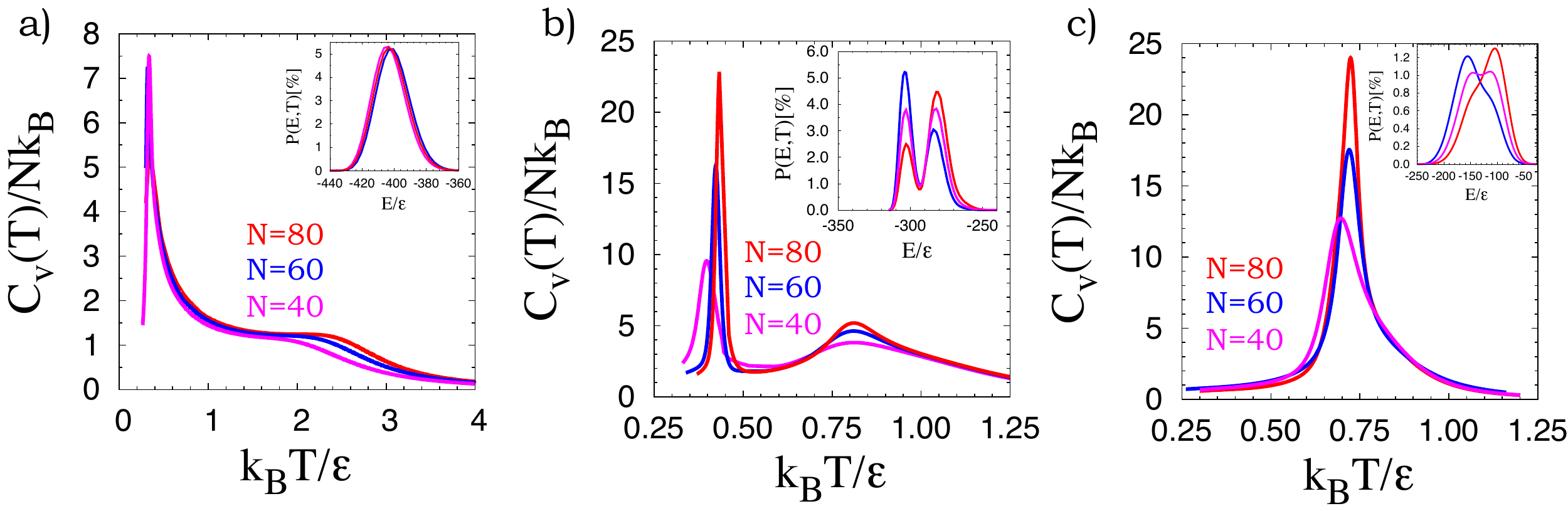}
\caption{Specific heat per bead ($C_v/Nk_B$) as a function of reduced
temperature ($T^*$=$k_B T/\varepsilon$) for chains of length $N=$ 40, 60, and 80. Panel a) corresponds to $R_c=1.6$ and $\sigma_{SC}/\sigma = 0$, for which the ``ground state" is an unstructured compact globule. The specific heat curve suggests the existence of two phase transitions. The transition at the higher temperature, between the coil and the collapsed globule (the $\theta$-transition), is signalled by the shoulder. The more pronounced lower temperature peak is suggestive of a second continuous transition between two unstructured phases, the collapsed globule and the compact globule. Panel b) corresponds to $R_c = 1.6$ and $\sigma_{SC}/\sigma = 1.0$ -- the ground state is the (7,45) precessing helix (shown in panel 3 of Figure 3). Panel c) shows the specific heat curve for $R_c = 1.6$ and $\sigma_{SC}/\sigma = 1.5$ -- the ground state here is the (1,5) dual helix (shown in panel 4 of Figure 3). In all cases, the inset shows the canonical probability distribution of the energy in the vicinity of the transition to the ground state (the lower transition temperature when there are two transitions) for the longest chain studied ($N=80$) at three temperatures: the transition temperature corresponding to the peak in the specific heat (purple curve), a temperature 1\% below the transition temperature (blue curve), and a temperature 1\% above the transition temperature (red curve). The numbers of attractive contacts in the lowest energy conformations in our simulations are 443, 316, and 259 for $\sigma_{SC}/\sigma = $ 0, 1, and 1.5 respectively. The key message is that the low temperature transition in the absence of side spheres appears to be continuous, whereas the addition of large enough side spheres results in a structured helical ground state and a first order transition to it upon lowering the temperature. The low-temperature first order transition weakens upon increasing the side sphere size (see panel c) versus b)) because of the decrease in conformational entropy in the restricted coil phase.
\label{fig:Fig_2}}
\end{figure}

\newpage
\begin{figure}[htpb]
\centering
\captionsetup{justification=raggedright,width=\linewidth}
\includegraphics[width=0.7\linewidth]{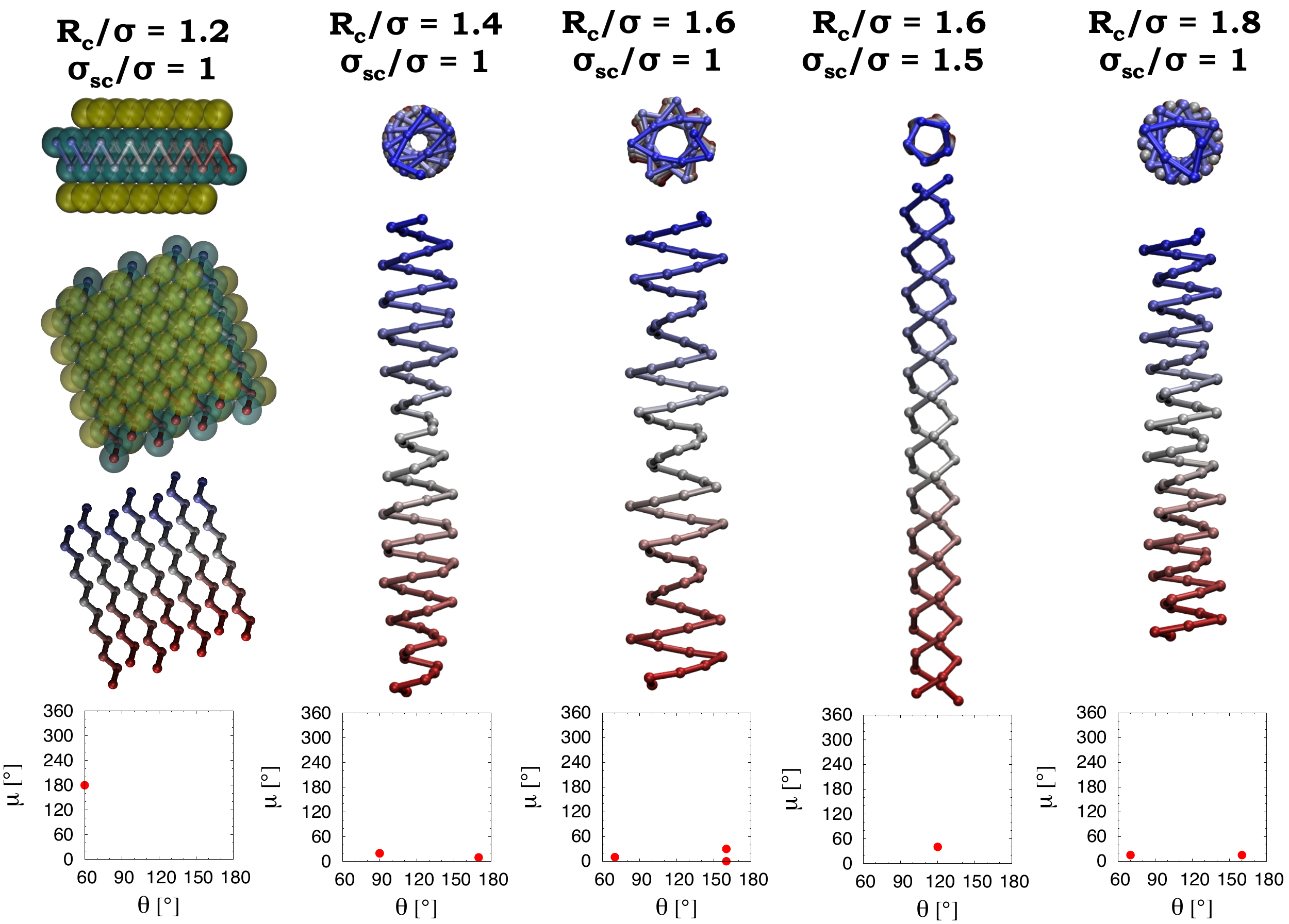}
\caption{Sketches of several ordered structures in the marginally compact phase of a tangent chain for different attraction ranges and side chain diameters. The structures are mathematical idealizations of those observed in the simulations having the same number of attractive contacts. Small variations around the idealized structures are permitted without any change in energy (the number of contacts). The helical structures shown are robust and reproducible over a range of chain lengths with two different simulation techniques and are likely ground states in the thermodynamic limit. The sheet structure is not observed in our simulations of a short chain but is mathematically constructed to maximize the number of contacts for the parameters shown. The side spheres are explicitly shown only for the sheet. For the helices, the side spheres stick out tangentially. In all cases, there are no steric clashes. The ground state for $R_c/\sigma = 1.2$ is a (1,2) sheet with 6 contacts per interior sphere and a repeat of ($\theta$,$\mu$) values of ($60^\circ$,$180^\circ$); for $R_c/\sigma = 1.4$, a (7,39) helix with (5+5+8+...) contacts per main chain sphere (degenerate with the sheet and a dual helix) corresponding to repeat ($\theta$,$\mu$) values of ($90^\circ$,$20^\circ$), ($90^\circ$,$20^\circ$), and ($170^\circ$,$10^\circ$); for $R_c/\sigma = 1.6$ and $\sigma_{SC}/\sigma = 1.0$ a (7,45) helix with (6+10+10+...) contacts and a repeat of ($\theta$,$\mu$) values of ($60^\circ$,$180^\circ$), ($160^\circ$,$30^\circ$), and ($160^\circ$,$0^\circ$); for $R_c/\sigma = 1.6$ and $\sigma_{SC}/\sigma = 1.5$, a dual helix with 6 contacts per bulk main chain sphere and a repeat of ($\theta$,$\mu$) values of ($120^\circ$,$40^\circ$); and for $R_c/\sigma = 1.8$, a (7,38) helix with (12+16+...) contacts and a repeat of ($\theta$,$\mu$) values of ($70^\circ$,$15^\circ$) and ($160^\circ$,$15^\circ$). The notation used is (number of approximate full turns per period of the helix,number of main chain spheres per period). The sheet picture does not depict the turns connecting the zig-zag strands and the double helix picture does not show the single turn. 
\label{fig:Fig_3}}
\end{figure}

\begin{figure}[htpb]
\centering
\captionsetup{justification=raggedright,width=\linewidth}
\includegraphics[width=0.7 \linewidth]{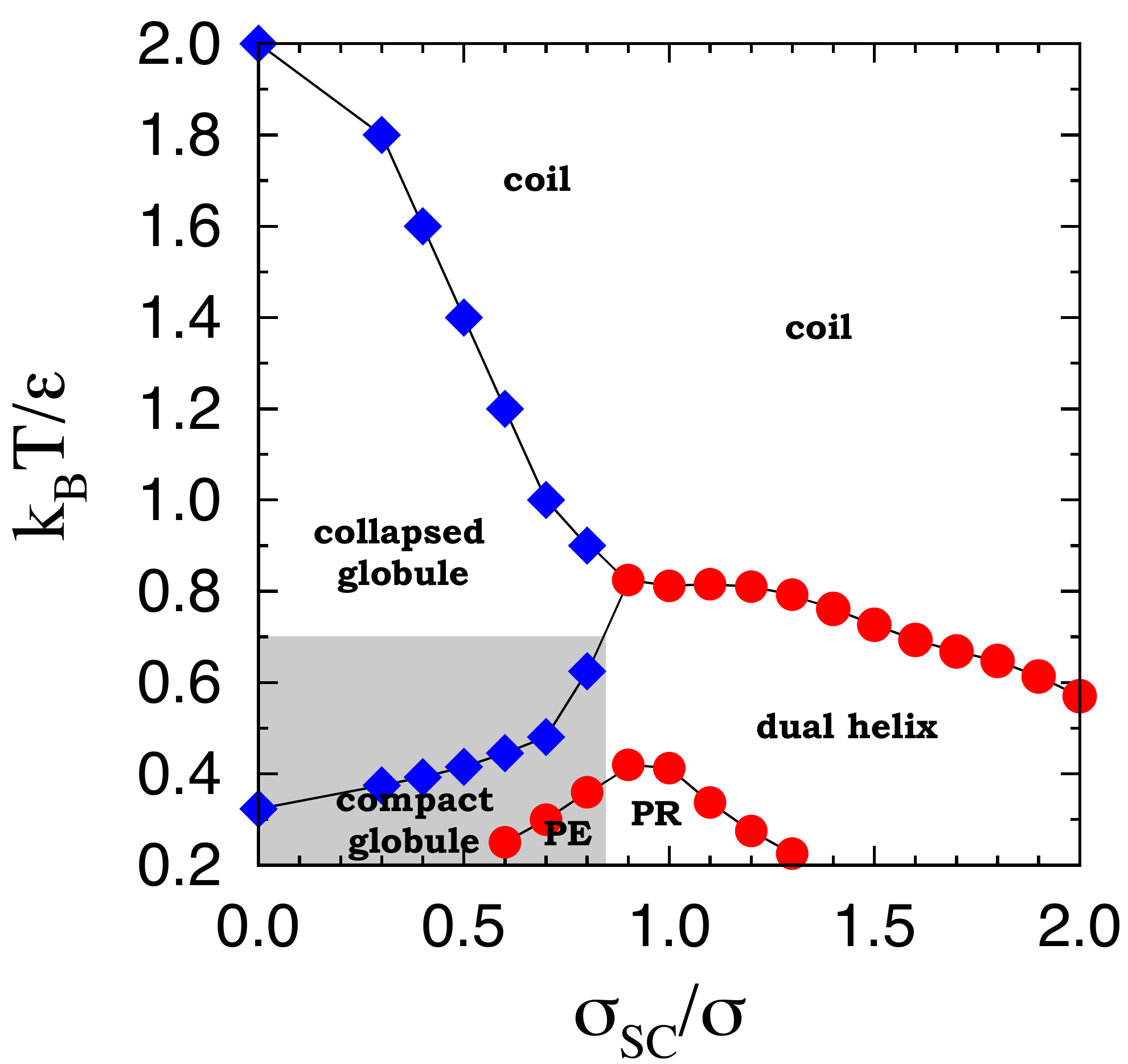}
\caption{Phase diagram in the reduced temperature -- side chain diameter plane for the tangent polymer chain of length $N=60$. The blue points (diamonds) indicate continuous transitions whereas the red points (circles) are first order transitions. Both parallel tempering and Wang-Landau simulations yield consistent results. The shaded region indicates a part of the phase diagram where the relatively modest chain length and potential equilibration issues make the results possibly unreliable. In the unshaded region, the error estimates are smaller than the size of the points. The coil is the canonical high temperature phase. For small side spheres, there are two globule phases, a collapsed globule at intermediate temperatures (with the familiar $\theta$-transition between the coil and the collapsed globule) as well as a distinct low temperature phase of a compact globule (with a second continuous transition between the compact and collapsed globule phases). The penetrated helix phase (PE) has, as its ground state, a single or a dual helix with a large enough radius to allow penetration of a rod (an essentially straight chain segment) within it (see Figure 5), the precessing helix phase (PR) has a helical ground state with a period of 45 main chain spheres yielding a rotation of approximately 7 turns (see third panel in Figure 3), and the dual helix is comprised of two symmetric helices connected by a turn (see fourth panel in Figure 3 -- turn not shown).
\label{fig:Fig_4}}
\end{figure}

\newpage

\begin{figure}[htpb]
\centering
\captionsetup{justification=raggedright,width=\linewidth}
\includegraphics[width=0.5 \linewidth]{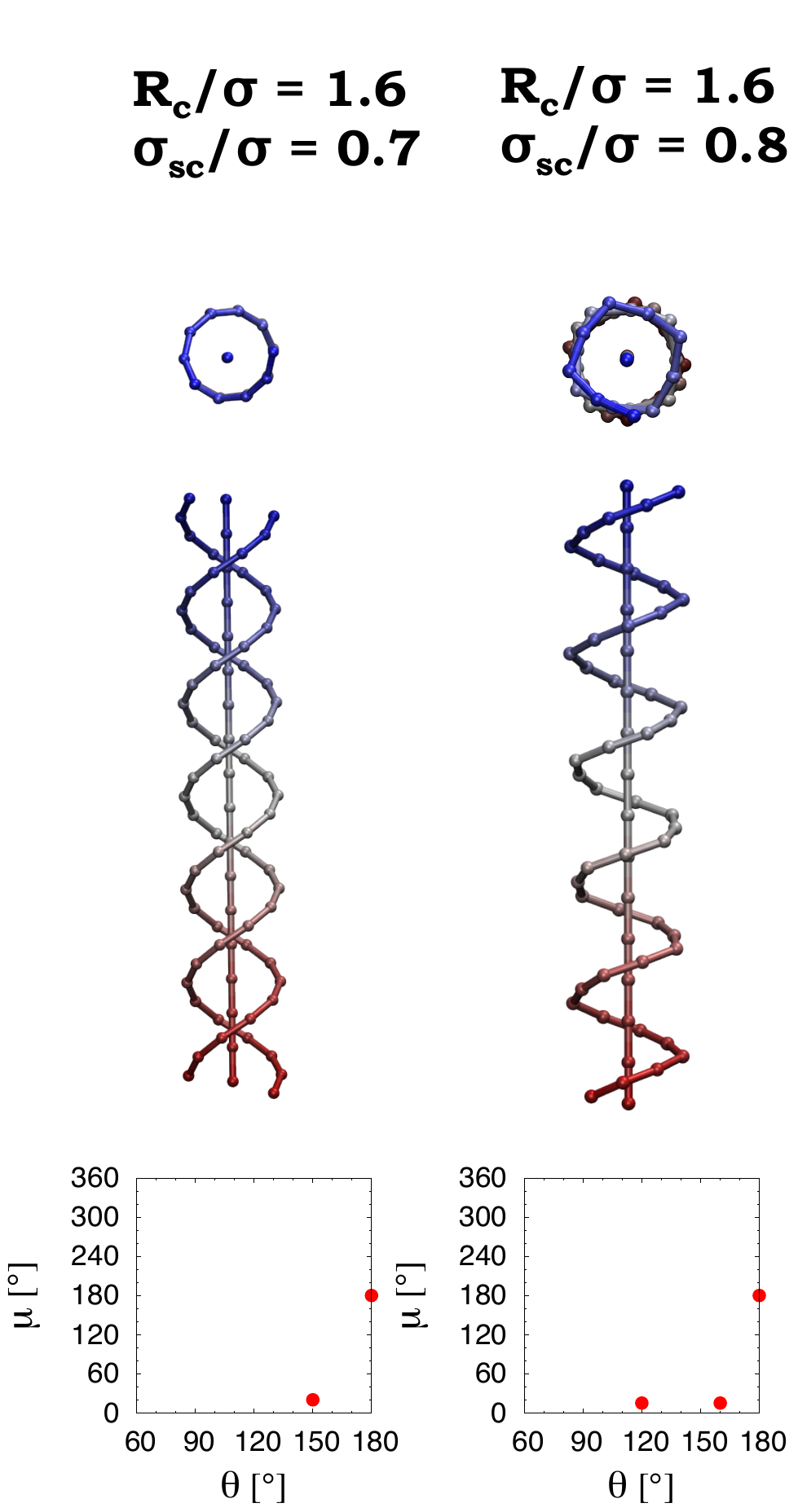}
\caption{Mathematical idealizations of two structures. The two panels show penetrated helical structures, corresponding to a single rod (an essentially straight chain segment) within a dual helix and a single helix. They are found for $R_c/\sigma = 1.6$ and for $\sigma_{SC}/\sigma = $0.7, and 0.8 respectively. The point corresponding to both $\theta$ and $\mu$ equal to $180^\circ$ denotes the rod whereas the small $\mu$ values correspond to the helical jackets. These structures are seen consistently in our computer simulations, they can be sustained in the long chain limit, and they are probably ground state structures.   
\label{fig:Fig_5}}
\end{figure}

\end{document}